\documentclass[12pt, a4paper]{article}
\usepackage[utf8]{inputenc}
\usepackage{geometry}
\usepackage{amsmath, amssymb, amsthm}
\usepackage{graphicx}
\usepackage{natbib}
\usepackage{hyperref}
\usepackage{titlesec}
\usepackage{booktabs}
\usepackage{float}

\newtheorem{proposition}{Proposition}
\newtheorem{assumption}{Assumption}

\title{Suppression and Deterrence: Revisiting the Welfare Consequence of HIV Stigma}
\author{Pengyu Li}
\date{}

\begin{document}

\maketitle

\begin{abstract}
The government's effort to alleviate HIV stigma has been justified by the suppression effect of stigma on the HIV testing rate. Nevertheless, the deterrence effect of stigma on undesirable sexual behaviours has long been overlooked. This study adapts the existing framework on HIV stigma with an additional stage that formally models people's choices on whether to take preventive measures in sex. The model shows that, when sex is explicitly modelled, the suppression and deterrence effect coexist, which makes the net societal impact of HIV stigma ambiguous. A utilitarian welfare analysis concludes that the welfare-maximizing stigma level can be higher than its natural level, implying that the government's effort to reduce stigma is not always welfare-improving. Instead, the study provides a rationale for maintaining a certain level of HIV stigma to maximize social welfare.
\end{abstract}

\section{Introduction}

Social stigma is the prejudice or exclusion an individual or a group faces in society due to some attributes or characteristics they possess that mark them as different (as cited in Major and O'Brien (2005)). It has both psychological consequences on the individual level, such as lowering self-esteem and impairing physical and mental health (Major and O'Brien, 2005) and societal consequences depending on its context. One salient example is HIV stigma, the type of stigma that is specifically against those who are diagnosed with, or at risk of, HIV/AIDS.

HIV stigma has been regarded as the main cause for the insufficient HIV testing rate, which is undesirable for timely initiating treatment and containing transmission (Yang et al., 2023). Derksen, Muula and van Oosterhout (2020) modelled stigma and testing using a Bayesian game. Their model shows that high stigma suppresses the testing rate because the choice of testing is perceived as a signal of high risk. Following this argument, governments have been taking proactive actions to tackle stigma. For example, the UK's "Towards Zero" action plan to end HIV in England (2022-2025) set addressing stigma as one of its key objectives (See Objective 4). The UK government is committed to spending on stigma-relieving programs, educating the public and improving knowledge and understanding across the healthcare system to tackle stigma (Department of Health and Social Care (UK), 2021).

Nevertheless, HIV stigma's potential role as a deterrence towards undesirable behaviour has been given little attention. Funk (2004) discussed how stigma against convicted criminal offenders acts as a crime-deterrent. His model shows that stigma, when applied appropriately, reduces work-related crimes and improves labour market efficiency. Analogously, since HIV typically results from inappropriate sexual practices, e.g., lack of adoption of HIV-preventive measures like condoms, stigma might be capable of correcting this behaviour. Unfortunately, this idea cannot be conveyed by the latest applicable models of HIV stigma (Derksen, Muula and van Oosterhout, 2020; Yang et al., 2023) since people's choices in sexual practices have not been explicitly modelled.

This study aims to fill in this gap by improving the model developed by Derksen, Muula and van Oosterhout (2020) and Yang et al. (2023) with an additional stage of formally modelled sexual behaviours. As a starting point, I identify two factors that jointly rationalize the non-universal use of preventive measures like condoms, an ex-post irrational decision. The first one is the cost of its usage. Having sex with condoms, for instance, requires time to prepare, money to purchase the product, and the experience is generally less pleasurable. Moreover, Aronson (2011) points out that condoms tend to be seen as unromantic and remind people of disease, which adds to the psychological burden for two people who are about to make love. To sum up, having sex with preventive measures generates less instantaneous pleasure than having sex without. The other factor is a lack of self-control in an emotionally intense state. A survey among college students by Ariely and Loewenstein (2006) revealed that, compared with subjects under normal states, sexually aroused subjects are more likely to claim that condoms "decreases sexual pleasure", "interferes with sexual spontaneity", and are less likely to use condoms if the sexual history of a new sexual partner is unknown.

It seems reasonable to conjecture that emotional impulse overrides rational objection against unsafe sex while people are in sexual arousal. Regarding its mechanism, prior literature on decision-making in sexual coercion found that sexual arousal gives rise to the overperception of sexual intent (Bouffard and Miller, 2014) and the instantaneous importance of sexual pleasure (Bouffard, 2002). Combining these two arguments, I formally model the choice of unsafe sex as a self-control issue, where people with strong present bias overweigh its immediate benefit relative to subsequent consequences.

The theoretical model is a two-period game. In the first period of the game, pairs of risky individuals play coordination games with sufficient pre-play communication. The heterogeneity in present bias drives coordination on different equilibria, which endogenously determines players' risk types. In the second period, I model stigma against HIV testing as a signalling Bayesian game where players take their risk types as private information. The model reveals that, when both sex and testing choices are formally modelled, the suppression and deterrence effects of stigma coexist, and its societal impact becomes ambiguous.

To quantitatively analyze the overall social impact and propose policy implications, I construct a utilitarian welfare analysis. Similarly to O'Donoghue and Rabin (2006), I treat the present bias, the cause of unsafe sex choices, as a behavioural error that deviates "decision utility" from "experience utility". Welfare calculation is based on the latter, where the importance of the future is not underweighted. The analysis shows that government spending in reducing stigma to its natural level is not always welfare-improving. When the deterrence effect dominates the suppression effect, further alleviating HIV stigma improves the testing rate at a higher cost of increasing the fraction of unsafe sex and people with high HIV risk. The model shows that a welfare-maximizing stigma level is theoretically possible to be higher than its natural level.

The remainder of this study is structured as follows: Section 2 presents the theoretical framework of HIV stigma. Then, Section 3 provides a welfare analysis with a numerical example. Section 4 concludes the study.

\section{Theoretical Model}

The context of HIV is modelled as a two-period game. In the first period, rational individuals from population $\mathcal{A}$ are randomly matched with each other in pairs. Each pair of individuals plays a coordination game with sufficient pre-play communication by choosing to engage in safe or unsafe sex, which determines their HIV risk type in the second period. The second period comprises two stages. Individuals with privately known HIV risk types decide whether to receive HIV tests in the first stage, where test results are not revealed but testing choices are publicly observable. Then, in the second stage, people in another population $\mathcal{B}$ are matched one-on-one with population $\mathcal{A}$ and choose to accept or reject social interaction contingent on testing choices. I proceed with the discussion on the model via backward induction.

\subsection{Period 2. Testing (Stage 2) and Interaction (Stage 3)}

The model of period 2 is closely based on that of Yang et al. (2023) and Derksen, Muula and van Oosterhout (2020), who model the suppressing effect of HIV stigma on testing. I adapt their models with modifications for compatibility with the newly added period 1 and provide additional intuition.

\subsubsection{Environment}
Consider a pair of two players, $A$ and $B$, each of which is randomly drawn from two continua of populations, $\mathcal{A}=[0,1]$ and $\mathcal{B}=[0,1]$. Each player $A$ has a type $(\theta_{a}, y_{a})$, where $\theta_{a} \in \{\theta_{L}, \theta_{H}\}$ represents the possibility of $A$ being HIV-positive with $0 < \theta_{L} < \theta_{H} < 1$ and $y_{a}$ is the private valuation of social interaction in stage 3. The risk type, $\theta_{a}$, of each player $A$ from population $\mathcal{A}$ is determined by his choice in period 1 and is taken as given by player $A$ in period 2. Use $r$ to denote the endogenously determined fraction of $A$ with $\theta_{a}=\theta_{H}$. Denote the true HIV status as $h_{a} \in \{0 \text{ HIV negative}, 1 \text{ HIV positive}\}$, which is never observed by player $B$, and the fraction of HIV-positive player $A$ as $\bar{h}$. Each player $B$ also has $(\theta_{b}, y_{b})$ where $\theta_{b}=0$ is publicly known and $y_{b}$ is privately known. $y_{a}$ and $y_{b}$ are independent and identically distributed (i.i.d.) with CDF $\mathcal{G}$ on support $(\epsilon, +\infty)$ with $\epsilon > 0$ being sufficiently small\footnote{The condition $\epsilon>0$ is only important for the existence of the pooling equilibrium (see Equation A.2, Online Appendix, Yang et al., 2023), which nevertheless will be dismissed from application later. Setting $\epsilon=0$ has no consequence on the validity of the partially separating equilibrium.}, which represents mutual gains from all social interactions like business relations, employment, and sex. $\bar{h}, \mathcal{G}, r$ are common knowledge.

In stage 2, $t_{a} \in \{0 \text{ not test}, 1 \text{ test}\}$ denotes player $A$'s choices on whether to take an HIV test, which is publicly observable to all players. If choosing $t_{a}=0$, player $A$ receives a net payoff of 0 in this stage. If choosing $t_{a}=1$, player $A$ bears a direct cost of $c$, which summarizes all the time, monetary, physical, and mental costs associated with the action of taking the test. If the test result returns positive, player $A$ is assumed to receive treatment immediately with a health benefit of $v$. Moreover, $h_{a}=1$ (HIV infection) is associated with an additional cost of $c_{h}$ (a value satisfying Assumption~\ref{ass:utility-gap}), comprising all the physiological and psychological costs of being infected with HIV, which is not dependent on testing choices.

In stage 3, upon observing $t_{a}$ of the randomly matched player $A$, each player $B$ chooses $m_{b} \in \{0 \text{ not interact}, 1 \text{ interact}\}$. Given choice of $m_{b}=0$ (social exclusion), both players receive a net payoff of 0. Given the choice of $m_{b}=1$ (social inclusion), both player $A$ and $B$ receive their valuation $y_{a}$ and $y_{b}$, respectively, while player $B$ now bears a cost of potential HIV transmission. The perceived HIV transmission probability is $\hat{\tau} \in [0,1]$\footnote{Similar to Derksen, Muula and van Oosterhout (2020), I make no assumption on the relationship between the perceived and true probability of transmission. The relationship is re-introduced in welfare analysis.}, which is homogeneous across all players in $\mathcal{B}$, and a health cost of $z$ once infected. Player $A$'s consent is not explicitly modelled since he always receives net gain from social interaction.

\subsubsection{Strategy and Equilibria}
In stage 2, player $A$'s payoff is given by:
\begin{equation*}
    u_a^{2}(t_a) = t_a(h_a v - c) - h_a c_h
\end{equation*}
His expected payoff of taking and not taking an HIV test is thus given, respectively, by:
\begin{align*}
    E_{h_{a}}[u_{a}^{2}(t_{a}=1)] &= E(h_{a})v - c - E(h_{a})c_{h} = \theta_{a}(v-c_{h}) - c \\
    E_{h_{a}}[u_{a}^{2}(t_{a}=0)] &= -E(h_{a})c_{h} = -\theta_{a}c_{h}
\end{align*}
Hence, the net gain of testing for player $A$ is:
\begin{equation*}
    E_{h_{a}}[u_{a}^{2}(t_{a}=1) - u_{a}^{2}(t_{a}=0)] = \theta_{a}v - c
\end{equation*}
and the additional cost of infection $c_{h}$ is irrelevant to the testing choice.

In stage 3, given matched player $B$'s choice $m_{b}$, player $A$'s payoff is:
\begin{equation*}
    u_a^{3}(m_b) = m_b y_a
\end{equation*}
The expected payoff of player $B$ herself, given her choice $m_{b}$ and belief on player $A$'s risk type $\hat{\theta}_{b}(t_{a})$, is given by:
\begin{equation*}
    u_{b}(m_{b}) = m_{b}[y_{b} - \hat{\tau}\hat{\theta}_{b}(t_{a})z]
\end{equation*}
The following two assumptions are introduced:

\begin{assumption}[Participation of Testing]\label{ass:par-testing}
\begin{equation*}
    \theta_L v < c < \theta_H v
\end{equation*}
High (low) risk individuals obtain positive (negative) net gain from testing alone since their risk of HIV infection is high (low).
\end{assumption}

\begin{assumption}[Participation of Social Interaction]\label{ass:par-interact}
\begin{equation*}
    P(y_b > \hat{\tau} \bar{h} z) = 1
\end{equation*}
Player $B$ will consent to social interaction with a player $A$ randomly drawn from $\mathcal{A}$.
\end{assumption}
Note that this assumption implies that $\mathbb{P}(y_{b} > \hat{\tau}\theta_{L}z)=1$, i.e., player $B$ never reject a known low-risk player $A$, since $\theta_{L} < \bar{h} < \theta_{H}$ in population $\mathcal{A}$.

\begin{proposition}[Equilibrium]
This game has two pure-strategy Perfect Bayesian Equilibria, a partially separating equilibrium and a pooling equilibrium.
\end{proposition}

\paragraph{The partially separating equilibrium}
In this equilibrium, a fraction $S$ of players in $\mathcal{B}$ rejects if and only if matched player $A$ took the test, while the remaining fraction accepts regardless. Taking this strategy and $S$ as given, player $A$ solves the following problem to maximize expected utility:
\begin{equation*}
    \max_{t_a \in \{0,1\}} t_a(\theta_a v - c) - \theta_a c_h + (1 - t_a S)y_a = t_a(\theta_a v - c - Sy_a) + (y_a - \theta_a c_h)
\end{equation*}
Hence, his best response is:
\begin{equation*}
    BR_{t_{a}}(S,y_{a}) = \begin{cases} 
    1, & \theta_{a}v - c - Sy_{a} > 0 \quad (\text{or equivalently, } y_{a} < \frac{\theta_{a}v-c}{S}) \\ 
    0, & \text{otherwise}
    \end{cases}
\end{equation*}
Intuitively, player $A$ takes the test if the health benefits of testing $(\theta_{a}v-c)$ is high enough or the valuation of social interaction $(y_{a})$ is low enough so that he is willing to face the risk of social exclusion against testing. Given Assumption~\ref{ass:par-testing}, all low-risk and a fraction of high-risk individuals from $\mathcal{A}$ with lower valuation $y_{a}$ never take the test. Only some high-risk individuals seek HIV tests, which implies adverse selection. With Bayesian updating, player $B$ forms the following consistent belief:
\begin{equation*}
    \hat{\theta}_{b}(t_{a}=1)=\theta_{H}, \quad \hat{\theta}_{b}(t_{a}=0)\in(\theta_{L},\bar{h})
\end{equation*}
Taking player $A$'s strategy and belief $\hat{\theta}_{b}(t_{a})$ as given, player $B$ solves the following problem to maximize expected utility:
\begin{equation*}
    \max_{m_b \in \{0,1\}} m_b [y_b - \hat{\tau}\hat{\theta}_b(t_a)z]
\end{equation*}
By Assumption~\ref{ass:par-interact}, player $B$'s best response is given by:
\begin{equation*}
    BR_{m_{b}}(t_{a},y_{b}) = \begin{cases} 
    1, & t_{a}=0 \\ 
    1, & t_{a}=1 \text{ and } y_{b} > \hat{\tau}\theta_{H}z \\ 
    0, & \text{otherwise}
    \end{cases}
\end{equation*}

An alternative representation of $BR_{m_{b}}(t_{a},y_{b})$ is shown in the table below:

\begin{table}[H]
\centering
\begin{tabular}{lccc}
\toprule
$BR_{m_{b}}(t_{a},y_{b})=?$ & $t_{a}=0$ & $t_{a}=1$ & \\
\midrule
$y_{b} > \hat{\tau}\theta_{H}z$ & 1 & 1 & $\rightarrow 1-S$ \\
$y_{b} < \hat{\tau}\theta_{H}z$ & 1 & 0 & $\rightarrow S$ \\
\bottomrule
\end{tabular}
\end{table}

In this equilibrium, $S$, the fraction of healthy people who reject social interaction if and only if observing their matches took an HIV test before, acts as an index for HIV stigmatization. It can be calculated using $\mathcal{G}$ as:
\begin{equation*}
    S = \int_{\epsilon}^{\hat{\tau}\theta_{H}z} d\mathcal{G}
\end{equation*}
The total testing rate among the population $\mathcal{A}$ and the testing rate among high-risk individuals are given, respectively, by:
\begin{equation*}
    R = r\int_{\epsilon}^{\frac{\theta_{H}v-c}{S}} d\mathcal{G}; \quad R_{H} = \frac{R}{r} = \int_{\epsilon}^{\frac{\theta_{H}v-c}{S}} d\mathcal{G}
\end{equation*}
These results lead to Proposition~\ref{prop:suppression}.

\begin{proposition}[Suppression Effect]\label{prop:suppression}
An increase (decrease) in perceived HIV transmission risk $\hat{\tau}$ leads to a higher (lower) level of HIV stigmatization $S$, which in turn results in a lower (higher) testing rate $R$ and $R_{H}$ among the population.
\end{proposition}

\paragraph{The pooling equilibrium}
In the pooling equilibrium, a proportion $P$ of player $B$ discriminates against non-testing. Hence, player $A$ solves the following problem to maximize expected utility:
\begin{equation*}
    \max_{t_{a}\in\{0,1\}} t_{a}(\theta_{a}v-c) - \theta_{a}c_{h} + [1-(1-t_{a})P]y_{a}
\end{equation*}
\begin{equation*}
    = t_{a}(\theta_{a}v-c+Py_{a}) + [(1-P)y_{a}-\theta_{a}c_{h}]
\end{equation*}
High-risk player $A$ always tests since it yields both health benefits from testing $(t_{a}(\theta_{H}v-c))$ and social interaction $(t_{a}y_{a})$. By appropriately setting the off-equilibrium belief $\hat{\theta}_{b}(t_{a}=0)$, there exists a sufficiently high $P$ that drives all low-risk player $A$ towards testing and thereby sustain the pooling equilibrium.\footnote{For a more formal proof of the pooling equilibrium and the absence of other equilibrium, see Yang et al. (2023) and Derksen, Muula and van Oosterhout (2020).}

The pooling equilibrium is of less empirical value. The outcome of universal testing contradicts the empirical observation. Also, it violates the D1 criterion of Perfect Bayesian Equilibrium and is thus unstable (Derksen, Muula and van Oosterhout, 2020). Therefore, I assume the partially separating equilibrium is played out throughout the remainder of this paper.

\subsubsection{Continuation value}
In connection with period 1, the payoff of each player $A$ entering period 2 as a high- or low-risk type is calculated as:
\begin{equation*}
    V_{L}(y_{a}) = y_{a} - \theta_{L}c_{h}
\end{equation*}
\begin{equation*}
    V_H(y_a) = \begin{cases}
    y_a - \theta_H c_h & y_a > \frac{\theta_H v - c}{\int_{\epsilon}^{\hat{\tau}\theta_H z} d \mathcal{G}} \\
    y_a - \theta_H c_h + (\theta_H v - c - y_a \int_{\epsilon}^{\hat{\tau}\theta_H z} d \mathcal{G}) & \text{otherwise}
    \end{cases}
\end{equation*}
It is plausible to assume that rational individuals should always prefer to possess low risk at the beginning of period 2. Consequently, I assume the continuation value of low-risk individuals always exceeds that of high-risk individuals, regardless of valuation on social interactions and stigma, or:
\begin{equation*}
    V_{L}(y_{a}) - V_{H}(y_{a}) > 0, \quad \forall y_{a} \in (\epsilon, +\infty)
\end{equation*}
Given that $\frac{\partial V_{L}(y_{a})}{\partial y_{a}} \ge \frac{\partial V_{H}(y_{a})}{\partial y_{a}} > 0$, this condition is equivalent to Assumption~\ref{ass:utility-gap} below.

\begin{assumption}[High-low Utility Gap]\label{ass:utility-gap}
\begin{equation*}
    V_{L}(\epsilon) - V_{H}(\epsilon) = (\theta_{H} - \theta_{L})c_{h} - [(\theta_{H}v - c) - \epsilon \int_{\epsilon}^{\hat{\tau}\theta_{H}z} d\mathcal{G}] > 0
\end{equation*}
or equivalently,
\begin{equation*}
    c_{h} > \frac{(\theta_{H}v - c) - \epsilon \int_{\epsilon}^{\hat{\tau}\theta_{H}z} d\mathcal{G}}{(\theta_{H} - \theta_{L})}
\end{equation*}
\end{assumption}

\begin{figure}[H]
    \centering
    \includegraphics[width=0.75\textwidth]{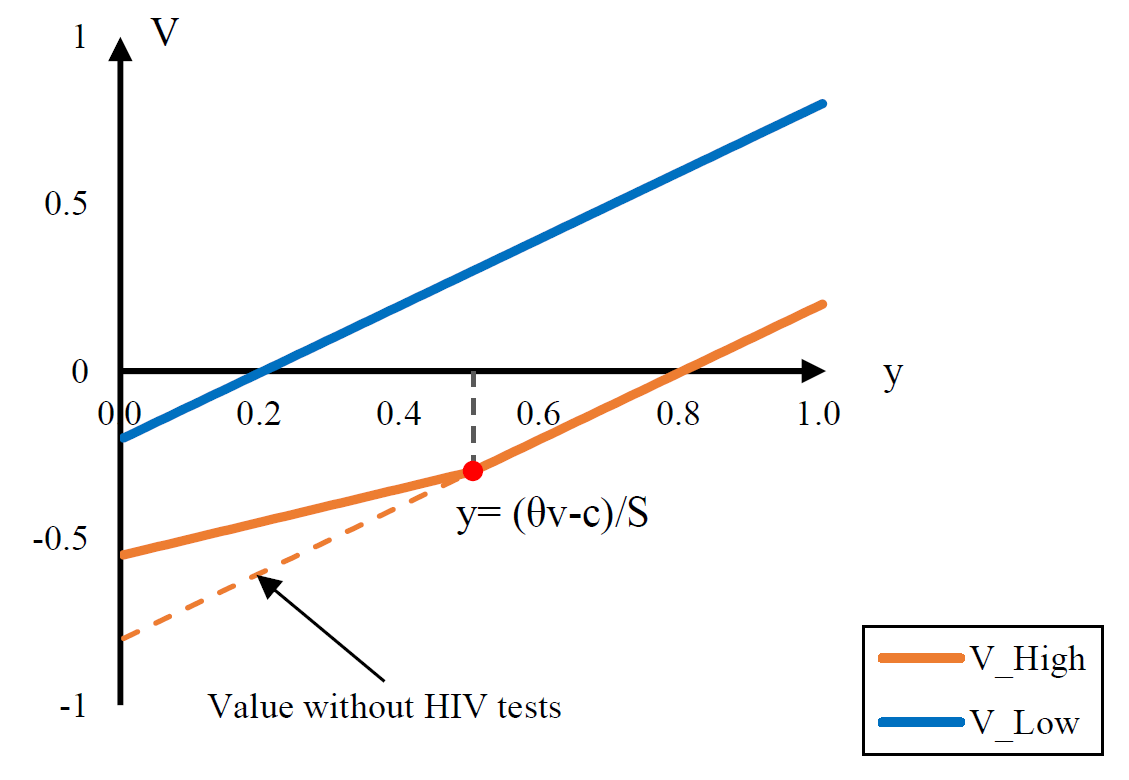}
    \caption{Graphical Illustration of Assumption~\ref{ass:utility-gap}}
    \label{fig:continuation_value}
\end{figure}

Essentially, this assumption demands a sufficiently large value of $c_{h}$. Note that $c_{h}$, a choice-irrelevant parameter, can be arbitrarily chosen and has no impact on equilibrium outcomes. A graphical illustrative example is given in Figure \ref{fig:continuation_value}.

This illustration is drawn based on the following parameter values: $\theta_{L}=0.2, \theta_{H}=0.8, S=0.5, v=1, c=0.55$. Given these parameter values, $c_{h}=1$ is sufficiently high to satisfy Assumption~\ref{ass:utility-gap} and thus adopted.

\subsection{Period 1. Sexual Behavior (Stage 1)}

The first stage of this game describes people's sexual behaviour as a two-player coordination game, where each of them chooses between conducting safe or unsafe sex. Those choices subsequently influence their risk type entering period 2. In a practical sense, safe sex involves the adoption of preventive measures like condoms, which reduces the HIV transmission risk considerably.

\subsubsection{Environment}
All player $A$ in population $\mathcal{A}=[0,1]$ are randomly matched in pairs at the beginning of period 1. Each pair of players will play a two-player complete-information coordination game with choice set $t_{sex} \in \{0. \text{unsafe}, 1. \text{safe}\}$. Only when the choices of two players "match", i.e., either (0,0) or (1,1), do both receive the same positive payoff from this period. If a match does not occur, both players receive infinitely large negative payoffs. The normal form is shown below:

\begin{table}[H]
\centering
\begin{tabular}{lcc}
\toprule
& Unsafe & Safe \\
\midrule
Unsafe & $(M, M)$ & $(-\infty, -\infty)$ \\
Safe & $(-\infty, -\infty)$ & $(M-u, M-u)$ \\
\bottomrule
\end{tabular}
\end{table}

where $M>0$ is some positive payoff from successful coordination and $u>0$ is the difference in payoff between coordinating at different equilibria. This difference in payoff is to reflect all additional costs of taking preventive measures that reduce HIV transmission risk.

The current model explicitly incorporates the impairment in rational decision-making under sexual arousal as heterogeneity in present bias. The utility of player $A$ in period 1 can be expressed as:
\begin{equation*}
    U_{A,1} = u_{a}^{1} + \beta\delta E_{1}(u_{a}^{2} + u_{a}^{3})
\end{equation*}
where $\beta \in (0,1)$ represents how one values the payoff in period 2 (testing and stigma) relative to the instantaneous payoff in period 1 (sex). I assume that $\beta$ follows CDF $\mathcal{F}$, where there is no gap or mass point on the support. For simplicity, the intertemporal discount factor $\delta$ is assumed to be 1 between periods 1 and 2.\footnote{For a game with only two periods, time-consistent $(\delta)$ and time-inconsistent $(\beta)$ discounting have equivalent effect on choices. Yet their distinction becomes clear later in welfare analysis, where the latter is considered as an error that deviates decision utility from experience utility, yet the former is not an error.}

The risk type of each player $A$, $\theta_{a}$, will be determined by the outcome of the coordination game. If a pair of players reach the "unsafe" match, i.e., (0,0), both players will inherit a high-risk type $(\theta_{H})$; otherwise, they will both inherit the low-risk type $(\theta_{L})$. The valuation of social interaction $y_{a}$ is not privately known until the beginning of period 2, but its distribution $\mathcal{G}$ is common knowledge throughout the game.\footnote{This is a crucial assumption. If $y_{a}$ has been privately known since the beginning of the game, it would surely influence choices in period 1. This greatly complicates the game since there are now two sources of heterogeneity ($\beta$ and $y$) that affects the equilibrium outcome. I make this assumption for the simplicity of the model.} Each player $A$'s choice in period 1 is not observed by players in $\mathcal{B}$, but $r$ and $\bar{h}$ become common knowledge once outcomes of all pairs in period 1 are determined.

Importantly, there is a stage of pre-game communication using "cheap talks", i.e., the exchange of messages that are not directly related to payoffs, before the coordination game. It allows for infinite rounds of negotiation until two players reach a consensus regarding the coordination. In the current modelling, I do not explicitly specify the strategic details of such communication. Instead, I draw results from various solution concepts in prior literature, while introducing plausible ad-hoc assumptions where existing solution concepts fall short of concluding a deterministic outcome.

\subsubsection{Payoffs and Equilibrium Selection}
The expected continuation value of entering period 2 as high- and low-risk type is given by:
\begin{equation*}
    EV_{L} = \int_{\epsilon}^{+\infty} V_{L}(y_{a})d\mathcal{G} = \int_{\epsilon}^{+\infty} y_{a}d\mathcal{G} - \theta_{L}c_{h}
\end{equation*}
\begin{equation*}
    EV_{H} = \int_{\epsilon}^{+\infty} V_{H}(y_{a})d\mathcal{G} = \int_{\epsilon}^{+\infty} y_{a}d\mathcal{G} - \theta_{H}c_{h} + \int_{\epsilon}^{\frac{\theta_{H}v-c}{\int_{\epsilon}^{\hat{\tau}\theta_{H}z}d\mathcal{G}}} [(\theta_{H}v-c) - y_{a}\int_{\epsilon}^{\hat{\tau}\theta_{H}z}d\mathcal{G}]d\mathcal{G}
\end{equation*}
Assumption~\ref{ass:utility-gap} implies that:
\begin{equation*}
    EV_{L} - EV_{H} = (\theta_{H} - \theta_{L})c_{h} - \int_{\epsilon}^{\frac{\theta_{H}v-c}{\int_{\epsilon}^{\hat{\tau}\theta_{H}z}d\mathcal{G}}} [(\theta_{H}v-c) - y_{a}\int_{\epsilon}^{\hat{\tau}\theta_{H}z}d\mathcal{G}]d\mathcal{G} > 0
\end{equation*}
To solve for players' period 1 strategy, I add the continuation value to the normal form with each payoff pair representing the expected utility for the entire game from the standpoint of period 1.

\begin{table}[H]
\centering
\begin{tabular}{lcc}
\toprule
& Unsafe & Safe \\
\midrule
Unsafe & $(M + \beta EV_H, M + \beta EV_H)$ & $(-\infty, -\infty)$ \\
Safe & $(-\infty, -\infty)$ & $(M-u+\beta EV_L, M-u+\beta EV_L)$ \\
\bottomrule
\end{tabular}
\end{table}

Given $\beta \in (0,1)$, all player $A$ can be categorized into two kinds. Those who have sufficiently strong present bias would prefer to coordinate on the "unsafe" equilibrium:
\begin{equation*}
    M + \beta EV_{H} > M - u + \beta EV_{L}
\end{equation*}
or equivalently,
\begin{equation*}
    0 < \beta < \frac{u}{EV_{L} - EV_{H}}
\end{equation*}
If $u > EV_{L} - EV_{H}$, this would apply to all player $A$. If $u < EV_{L} - EV_{H}$, then there will also be some who have milder present bias and prefer to coordinate on the "safe" equilibrium:
\begin{equation*}
    M + \beta EV_{H} < M - u + \beta EV_{L}
\end{equation*}
or equivalently,
\begin{equation*}
    \frac{u}{EV_{L} - EV_{H}} < \beta < 1
\end{equation*}
In the following discussion, I refer to these two kinds of people "\textit{hot}" (with strong present bias) and "\textit{cold}" (with mild present bias) individuals, respectively.

During the random pairing for population $\mathcal{A}$, two types of pairs emerge: pairs with similar levels of present bias (hot-hot or cold-cold) and pairs with distinct levels of present bias (hot-cold). Essentially, the two types differ in whether there is a conflict of interest in the coordination, which leads to different communication processes and equilibrium outcomes.

\paragraph{Similar present bias: Common-interest coordination}
When players with similar levels of present bias are paired with each other, both players in a coordination game prefer the same equilibrium, the unique Pareto-Efficient PSNE ("unsafe" equilibrium for hot-hot pairs and "safe" equilibrium for cold-cold pairs). Existing literature in pre-play communication made consensual propositions on the outcomes of such games. For example, Farrell (1987) shows that cheap talks improve the chances of coordination and, particularly, eliminate coordination failure in games with infinite rounds of communication and no conflict of interest. Rabin (1994) defines the concept of "negotiated equilibrium", which uniquely exists in the current game to be the efficient PSNE. His argument of preemptive concession applies: either player can silently consent to the other's preferred choice, which coincides with his favourite one. Furthermore, Kim and Sobel (1995) demonstrate that evolutionary stability also favours the efficient equilibrium in the same setting, which will be played by all in a population in the long term. In light of existing theories, I impose the following assumption without further detailing the communication stage.

\begin{assumption}\label{ass:hot-hot}
The unique Pareto-Efficient equilibrium is played for all hot-hot and cold-cold pairs.
\end{assumption}

\paragraph{Distinct present bias: Battle of the Sexes}
A conflict of interest arises in hot-cold pairs. There exist two PSNEs that are not Pareto-rankable: though both players want to have a match in sex choices, the cold person prefers the "safe" equilibrium, while the hot person prefers the "unsafe" equilibrium. Prior studies have not provided a solution concept that ranks these two PSNEs given communication, since both players can always "talk tough" and coordination failure persists even when communication rounds extend indefinitely (Farrell, 1987). Also, both equilibria, as well as any linear combination of them, are in the Pareto meet (his Definition 6) and satisfy the definition of "negotiated equilibrium" (Rabin, 1994). However, no ranking between them is proposed.

Indeed, Rabin (1994) acknowledges that his results merely provide a lower bound of what pre-play communication could achieve for coordination. In real life, natural language used in communication is highly sophisticated and effective in facilitating coordination (Crawford, 2016). On this basis, I impose the following ad-hoc assumption on outcomes for hot-cold pairs:

\begin{assumption}\label{ass:hot-cold}
The Pareto-Efficient equilibrium that maximizes the (expected) utilitarian welfare of two players is played for hot-cold pairs.\footnote{Formally, one can justify this assumption by incorporating social preferences (Charness and Rabin, 2002) and setting the altruism parameter $\rho=\sigma=0.5$:
\begin{equation*}
U_{A1,1}=U_{A2,1}=0.5u_{a1}^{1}+0.5u_{a2}^{1}+\delta E_{1}(\beta_{1}u_{a}^{2}+\beta_{2}u_{a}^{3})
\end{equation*}.
Although the magnitude of social preferences depends heavily on contextual factors (Levitt and List, 2007), one could still find this assumption plausible because a sexual relationship is more intimate and exclusive than relationships in other social contexts.}
\end{assumption}

Based on Assumption~\ref{ass:hot-cold}, both players should now prefer the same equilibrium that maximizes the utilitarian welfare of the two people. In other words, both players prefer the "unsafe" equilibrium if:
\begin{equation*}
    2(M-u) + \beta_{1}EV_{L} + \beta_{2}EV_{L} < 2M + \beta_{1}EV_{H} + \beta_{2}EV_{H}
\end{equation*}
where $\beta_{1}$ and $\beta_{2}$ denotes the present bias of player $A1$ and $A2$, respectively, or equivalently,
\begin{equation*}
    0 < \beta_{1} + \beta_{2} < \frac{2u}{EV_{L} - EV_{H}}
\end{equation*}
They prefer the "safe" equilibrium if otherwise. Essentially, Assumption~\ref{ass:hot-cold} reconciles the conflict of interest, imposing that the "unsafe" equilibrium is preferred if the present biases of a hot-cold pair are jointly stronger. Applying Assumption~\ref{ass:hot-hot} again, one concludes that players will coordinate on the equilibrium preferred by both.

\subsubsection{Equilibrium Path Outcomes}
The condition $u > EV_{L} - EV_{H}$ implies that "unsafe" sex is so much more pleasurable than "safe" sex that a person with no self-control problem would prefer it and bear a higher HIV risk. Under this condition, all pairs of player $A$ will play "unsafe" equilibrium in period 1 and inherit $\theta_{a} = \theta_{H}$. HIV testing loses all signalling value given common knowledge of $r=1$. There will be universal testing due to the positive net benefit of testing for high-risk people and social interaction is only accepted for pairs where player $B$ has $y_{b} > \hat{\tau}\theta_{H}z$. This case is less interesting because its implied equilibrium outcome drastically deviates from real-life data.

\begin{figure}[H]
    \centering
    \includegraphics[width=0.75\textwidth]{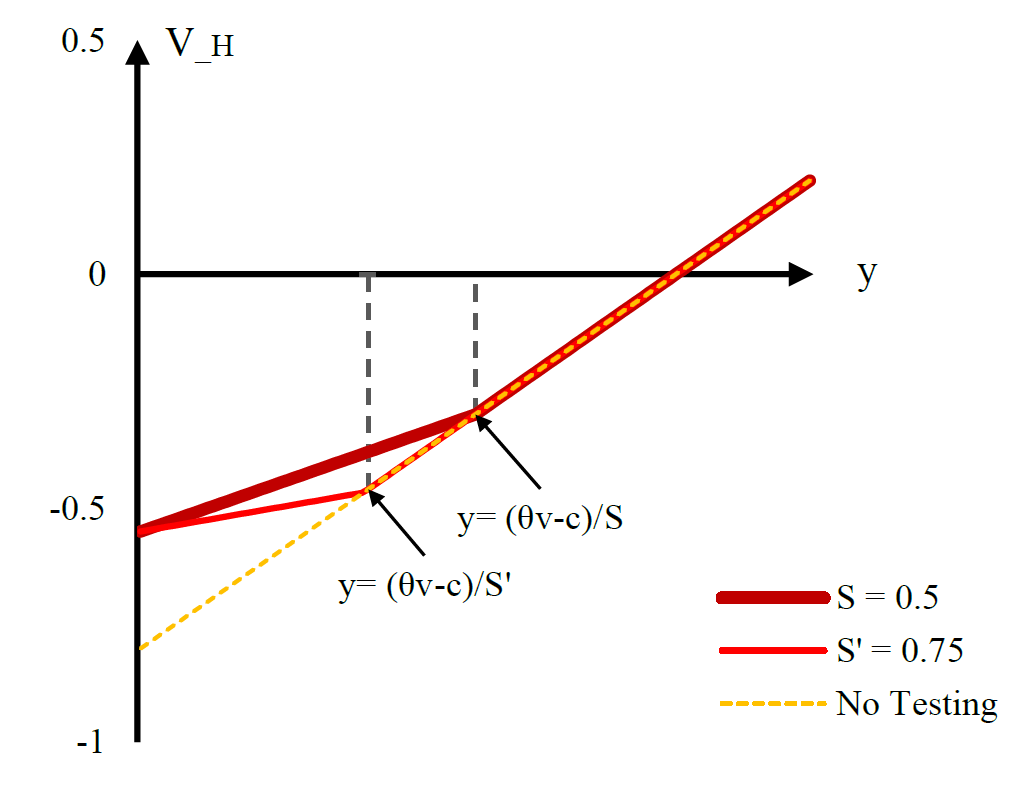}
    \caption{Effect of Increasing S on $V_{H}$}
    \label{fig:stigma_on_testing}
\end{figure}

Instead, I focus on the case of $u < EV_{L} - EV_{H}$. The proportion of "hot" people in $\mathcal{A}$, denoted by $H$, is given by:
\begin{equation*}
    H = \int_{0}^{\frac{u}{EV_{L} - EV_{H}}}d \mathcal{F}
\end{equation*}
where\footnote{Given that only partially separating equilibrium is considered, I make the simplification that $\epsilon=0$, i.e., $y$ is independent and identically distributed (i.i.d.) with CDF $\mathcal{G}$ on support $(0,+\infty)$.}
\begin{equation*}
    EV_{L} - EV_{H} = (\theta_{H} - \theta_{L})c_{h} - \int_{0}^{\frac{\theta_{H}v-c}{S}} [(\theta_{H}v-c) - y_{a}S]d\mathcal{G}
\end{equation*}
Note that $EV_{L} - EV_{H}$ increasing in $S$ since $EV_{H}$ is decreasing in $S$ (illustrated in Figure \ref{fig:stigma_on_testing}, using the same set of parameter values as \ref{fig:continuation_value}).

The same set of parameter values as Figure 1 is used for illustrative purposes.

\begin{figure}[H]
    \centering
    \includegraphics[width=1\textwidth]{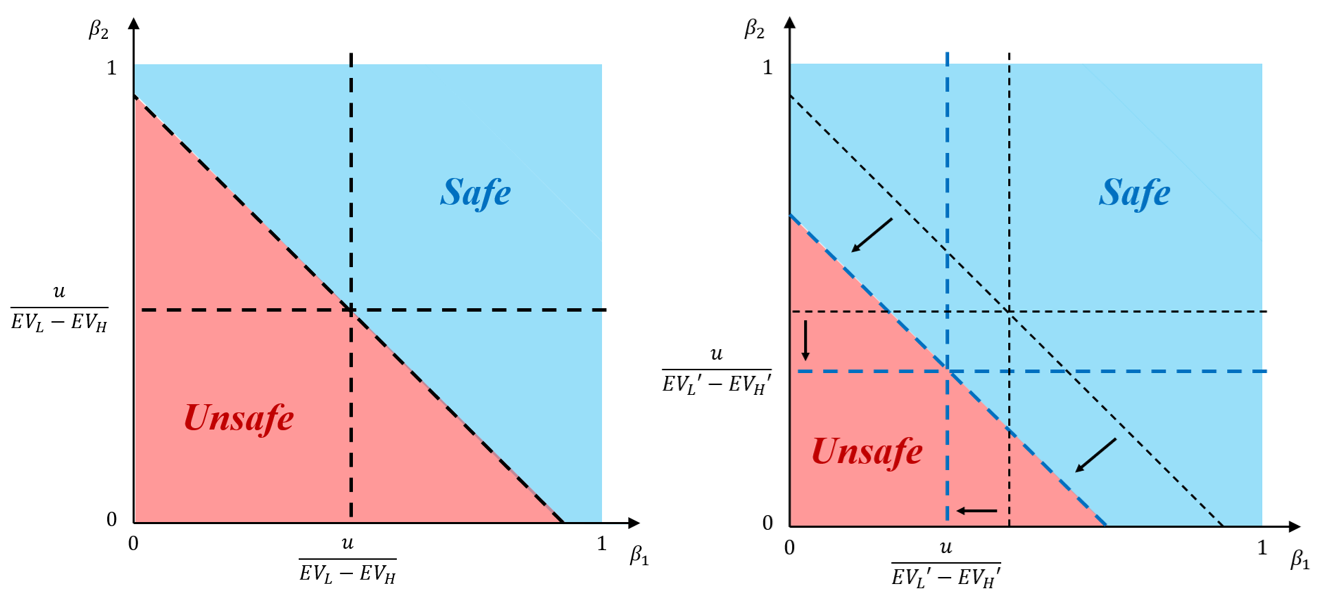}
    \caption{Effect of Increasing $EV_{L}-EV_{H}$ on $r$}
    \label{fig:continuation_on_sex}
\end{figure}

All hot-hot pairs $(H^2)$ will play "unsafe" equilibrium and all cold-cold pairs $((1-H)^2)$ will play the "safe" equilibrium. Among those hot-cold pairs, the proportion of pairs (among all pairs) coordinating on the "unsafe" equilibrium is given by:
\begin{equation*}
    2 \int_{\beta_{1}=\frac{u}{EV_{L}-EV_{H}}}^{\frac{2u}{EV_{L}-EV_{H}}} \left( \int_{\beta_{2}=0}^{\frac{2u}{EV_{L}-EV_{H}}-\beta_{1}} d\mathcal{F}(\beta_{2}) \right) d\mathcal{F}(\beta_{1})
\end{equation*}
The proportion of high-risk player $r$, which are those coordinated on "unsafe" equilibrium in period 1, is:
\begin{equation*}
    r = H^2 + 2 \int_{\beta_{1}=\frac{u}{EV_{L}-EV_{H}}}^{\frac{2u}{EV_{L}-EV_{H}}} \left( \int_{\beta_{2}=0}^{\frac{2u}{EV_{L}-EV_{H}}-\beta_{1}} d\mathcal{F}(\beta_{2}) \right) d\mathcal{F}(\beta_{1})
\end{equation*}
$r$ is decreasing in $EV_{L} - EV_{H}$ as graphically illustrated in Figure \ref{fig:continuation_on_sex}, where a widening in the gap $EV_{L} - EV_{H}$ shrinks the area of integration for "unsafe" equilibrium.

In period 2, given the signalling value of testing, the level of HIV stigma is proxied by the proportion $S$ of player $B$ who discriminates against testing. The effect of $\hat{\tau}$ on equilibrium outcomes in period 1 can be summarized as follow:

\begin{proposition}[Deterrence Effect]\label{prop:deterrence}
An increase (decrease) in perceived HIV transmission risk $\hat{\tau}$ leads to a higher (lower) level of HIV stigmatization $S$, which in turn results in more (less) coordination on the "safe" equilibrium and a decreased (increased) proportion of high-risk individuals $r$ in the population.
\end{proposition}

Intuitively, when the perceived transmission risk $\hat{\tau}$ increases, stigma $S$ increases since more player $B$ with not sufficiently high $y_{b}$ now starts to discriminate against player $A$ who received testing. It suppresses testing among player $A$ because the expected loss from a potential rejection of social interaction is now higher. It constitutes \textbf{the suppression effect} (Proposition~\ref{prop:suppression}). 

This larger probability of rejection from an increase in $S$ reduces the expected continuation value of high-risk type individuals into period 2, $EV_{H}$. It further widens the gap between the expected continuation value of different types, $EV_{L} - EV_{H}$, which becomes relatively more valuable when compared to the instantaneous payoff of sex weighted by present bias. Those who were previously on the margin, i.e., indifferent between two equilibria, now strictly prefer the "safe" equilibrium. It reduces the proportion of pairs playing the "unsafe" equilibrium, which lowers the fraction of high-risk people, $r$, entering period 2. This is \textbf{the deterrence effect} (Proposition~\ref{prop:deterrence}).

\section{Welfare Analysis}

The previous section introduces a model that captures both the suppression and deterrence effects of HIV stigma. Specifically, the suppression effect is unambiguously welfare-reducing. If the population $\mathcal{B}$ overestimates the true HIV transmission risk, i.e., $\hat{\tau} > \tau$, some player $B$ would turn down the interaction when it is beneficial to accept. The stigma $S(\hat{\tau})$ is higher than its natural level $S(\tau)$, which results in not only fewer players gaining from social interaction but also a suppression of HIV testing even though it generates positive net benefits. This insight is translated into the policy suggestion that actions lowering HIV stigma are welfare-improving when the transmission risk is overestimated (Derksen, Muula and van Oosterhout, 2020; Yang et al., 2023). However, the welfare consequence of stigma becomes ambiguous in the presence of the deterrence effect. This is because stigma encourages more safe sex and leads to a welfare-improving composition change in population $\mathcal{A}$, i.e., some people switching from high-risk to low-risk. Since this is what they would have chosen if they were not present-biased, high stigma corrects such errors and improves welfare. 

Now that there coexist two effects, the following analysis illustrates that government efforts in lowering stigma are not always welfare-improving when the transmission risk is overestimated.

\subsection{Welfare Calculation}
For simplicity, I assume that the true transmission risk of HIV in social interactions is $\tau=0$, which means that population $\mathcal{B}$ can only overestimate the transmission risk $(\hat{\tau} \ge \tau=0)$ in non-sexual social interaction. Under this parameter choice, player $B$ should always accept social interaction since there is no cost and the natural level is $S(\tau)=0$.

I adopt a utilitarian social welfare function, i.e., equal weights for all people from $A$ and $B$. Current social welfare consists of expected gains from sexual interaction and testing for all player $A$, and social interaction for all player $A$ and $B$. Similar to O'Donoghue and Rabin (2006), I treat the preference for immediate gratification as a behavioural error in self-control. Given $u < EV_{L} - EV_{H}$ and heterogeneity in $\beta$, people with low $\beta$ made an error by overvaluing the instantaneous payoff from sex relative to the subsequent consequence, which led to a choice that they would regret in hindsight. Echoing the idea in Kahneman (1994), I calculate welfare based on the "experience utility" where $\beta=1$ for all. Present bias is the only source of "decision utility" deviating from "experience utility" for player $A$.

The social welfare for $\mathcal{A}$ is the sum of the expected payoff for high-risk and low-risk people, weighted by the proportion of each type. The social welfare functions for $\mathcal{B}$ is the sum of the payoff for those who discriminate against testing $(y_{b} < \hat{\tau}\theta_{H}z)$ and those who do not $(y_{b} > \hat{\tau}\theta_{H}z)$. The probability of any player $B$ being matched with a tested player $A$ equals the testing rate $R$.

\begin{align*}
    W_{\mathcal{A}} &= r(M + EV_{H}) + (1-r)(M - u + EV_{L}) \\
    W_{\mathcal{B}} &= \int_{0}^{\hat{\tau}\theta_{H}z} R y_{b} d \mathcal{G} + \int_{\hat{\tau}\theta_{H}z}^{+\infty} y_{b} d \mathcal{G} \\
    W &= W_{\mathcal{A}} + W_{\mathcal{B}}
\end{align*}

\subsection{Implication}
In the following discussion, the policy instrument of interest is $\hat{\tau}$, player $B$'s perceived transmission risk. The model assumes this parameter can be actively influenced by government actions and spending, e.g., stigma-reliving programs.\footnote{For an example, see Derksen, Muula and van Oosterhout (2022).} Given other exogenous parameter values, this is equivalent to indirectly using stigma level $S$ as the policy instrument.

\subsubsection{Error-free first-best benchmark}
Regarding the first-best outcome, consider the case where $\beta=1$ for all player $A$ and $\hat{\tau} = \tau = 0$ for all player $B$. If the sexual behaviour (period 1) is not modelled, some $r \in (0,1)$ will be exogenously determined and taken by all players as given. In this case, $S(\tau) = S(\hat{\tau}) = 0$, which means there is no discrimination against HIV testing and social interaction will be accepted in all pairs. This is trivially optimal for player $B$ since they face no cost from social interaction $(\tau=0)$ and should always accept. This is also efficient for all player $A$ because, given no stigma, all players with positive net gain from testing, i.e., high-risk type, will do so. The testing rate is given by:
\begin{equation*}
    \lim_{S \to 0^+} R = \lim_{S \to 0^+} r \int_{0}^{\frac{\theta_H v - c}{S}} d \mathcal{G} = r, \quad \lim_{S \to 0^+} R_H = 1
\end{equation*}
Therefore, the welfare $W$ is a monotonically decreasing function of $\hat{\tau}$. If initially $\hat{\tau} > \tau=0$, the policy that lowers $\hat{\tau}$ is unambiguously welfare-improving by minimizing the suppression effect.

Upon the introduction of period 1, if $\beta=1$ for all player $A$, then all pairs will play "safe" sex equilibrium in period 1 due to $u < EV_{L} - EV_{H}$ by assumption and no high-risk individual would emerge. Now that $r=0$ is common knowledge, HIV testing again loses its signalling value: no player $A$ will choose testing in stage 2 $(R=0)$ and all interactions are accepted in stage 3. HIV stigma does not affect this outcome because the type it discriminates against is absent and any $\hat{\tau}$ would be efficient. Hence, the policy based on $\hat{\tau}$ has no implication, which potentially implies that maintaining \textit{any} status quo is just fine.\footnote{This result makes it challenging to construct marginal argument regarding the welfare consequence of varying $\hat{\tau}$ around the natural level of $S(\tau)$, i.e., $\frac{\partial W(S(\hat{\tau}))}{\partial \hat{\tau}}|_{\hat{\tau}=\tau}$.}
The welfare benchmark under this state is given by:
\begin{align*}
    W_{\mathcal{A}} &= M - u + EV_{L} = M - u + \int_{0}^{+\infty} y_{a}d \mathcal{G} - \theta_{L}c_{h} \\
    W_{\mathcal{B}} &= \int_{0}^{+\infty} y_{b}d \mathcal{G}
\end{align*}

\subsubsection{Welfare loss from present bias}
When people in $\mathcal{A}$ are present-biased to some extent, some of them would coordinate at the "unsafe" equilibrium. If players in $\mathcal{B}$ maintain perceived transmission risk at $\hat{\tau} = \tau = 0$ and hence never reject any interaction $(S=0)$, the error in self-control only causes a decrease in welfare for those who switched equilibrium, from $EV_{L}$ to $EV_{H}$. By Assumption~\ref{ass:utility-gap}, $EV_{L} > EV_{H}$ and the per-capita reduction in welfare is:
\begin{equation*}
    EV_{L} - EV_{H} = (\theta_{H} - \theta_{L})c_{h} - (\theta_{H}v - c) > 0
\end{equation*}
The proportion of high-risk people who experienced a reduction in welfare is $r$. Therefore, the total reduction in welfare is $\Delta W = r(EV_{L} - EV_{H})$.

\subsubsection{Policy implications using stigma}
Denote the policy $\hat{\tau} > \tau = 0$. In the new state of the world under the policy:
\begin{align*}
    & S(\hat{\tau}) = \int_{0}^{\hat{\tau}\theta_{H}z} d\mathcal{G} > S(\tau) = 0 \\
    & EV_{L}(\hat{\tau}) = EV_{L}(\tau) \\
    & EV_{H}(\hat{\tau}) < EV_{H}(\tau) \\
    & r(\hat{\tau}) < r(\tau) \\
    & R(\hat{\tau}) = r(\hat{\tau})\int_{0}^{\frac{\theta_{H}v-c}{S(\hat{\tau})}} d\mathcal{G} < R(\tau) = 1
\end{align*}
Due to the suppression effect, those who remain in the "unsafe" equilibrium are worse off because $EV_{H}$ decreases as the stigma against testing increases. Those among them who choose testing now face the risk of rejection, while those who avoid testing let go of the positive net gain from testing. The change (loss) in welfare for this fraction of player $A$ is:
\begin{equation*}
    r(\hat{\tau})(EV_{H}(\hat{\tau}) - EV_{H}(\tau))
\end{equation*}
Due to the deterrence effect, some pairs previously near the margin are nudged to coordinate on the "safe" equilibrium. Their gain from period 2 is strictly higher since switching to low-risk people increases utility in expectation (by Assumption~\ref{ass:utility-gap}). The change (gain) in welfare for these people is:
\begin{equation*}
    (r(\tau) - r(\hat{\tau}))(EV_{L} - EV_{H}(\hat{\tau}))
\end{equation*}
The policy also has undesirable consequences for people in $B$. A proportion $S$ of them now discriminate against testing due to $\hat{\tau} > 0$, which they should not do given $\tau=0$. When paired with a tested player $A$ (with probability $R$), they erroneously reject the interaction and forgo $y_{b}$. The change (loss) in welfare for these player $B$ is:
\begin{equation*}
    -SR * E(y_{b}|y_{b} < \hat{\tau}\theta_{H}z) = -SR * \int_{0}^{\hat{\tau}\theta_{H}z} \frac{y_{b}}{S} d\mathcal{G} = -R * \int_{0}^{\hat{\tau}\theta_{H}z} y_{b} d\mathcal{G}
\end{equation*}
Aggregating all the changes in welfare, the model shows that a policy of $\hat{\tau} > \tau = 0$ is welfare-improving from baseline if there exist $\hat{\tau} \in (0,1]$ such that:
\begin{equation*}
    W(S(\hat{\tau})) - W(S(\tau))
\end{equation*}
\begin{equation*}
    = (r(\tau) - r(\hat{\tau}))(EV_{L} - EV_{H}(\hat{\tau})) + r(\hat{\tau})(EV_{H}(\hat{\tau}) - EV_{H}(\tau)) - R\int_{0}^{\hat{\tau}\theta_{H}z} y_{b} d\mathcal{G} > 0
\end{equation*}
The desirability of the policy naturally depends on the parameter values and specific distributions applied in the model. It is worth noting that this policy does not constitute Pareto improvement since player $B$ and some player $A$ are strictly worse off, nor is Pareto improvement possible for $\hat{\tau}$-based policies. Essentially, the policy introduces an error (overestimation of transmission risk) to offset another (present bias) by inducing beneficial composition change while, in the meantime, hurting those who do not benefit.

\subsection{Numeric Example}
This section provides a numeric example by assigning numeric values and simple distributions to all model parameters. The calculation does not provide a general condition for the existence of a welfare-improving $\hat{\tau} > \tau$ yet preliminarily shows that it can easily be obtained with common parameter values.

For simplicity, I assume both the present bias parameter, $\beta$, and payoff from social interaction, $y$, follow uniform distribution. Specifically, I assume both $\beta$ and $y$ follow are uniformly distributed on $(0,1)$\footnote{One might argue that this assumption on the lower bound of the support of $\beta$ is too extreme. Admittedly, a more realistic assumption would be based on support (0.9, 1) (O'Donoghue and Rabin, 2006). However, since other model parameters are all based on imaginary payoff values without realistic reference, realism in $\beta$ does not render realism to the entire numeric example. Hence, I chose (0,1) purely for simplicity.} and $(0,2)$, respectively. Additionally, I assume the difference in the payoff of "safe" versus "unsafe" sex, $u=0.1$, and health cost for player $B$, $z=2.5$. Other parameter values are the same as the set of values adopted in Figure \ref{fig:continuation_value}: $\theta_{L}=0.2, \theta_{H}=0.8, v=1, c=0.55, c_{h}=1$. Note that this set of parameter values satisfies Assumption~\ref{ass:utility-gap} $(\epsilon=0)$:
\begin{equation*}
    c_{h} = 1 > \frac{(0.8 \times 1 - 0.55)}{(0.8 - 0.2)} = \frac{(\theta_{H}v - c)}{(\theta_{H} - \theta_{L})}
\end{equation*}
Since there is no readily available reference in prior literature regarding the determination of these values, I select them rather arbitrarily for illustrative purposes. No parameter value is set to be extreme. In particular, $y_{max}=2$ and $z=0.25$ are set to make perceived risk $\hat{\tau}$ and stigma $S$ numerically identical. It achieves the convenient result that a completely misinformed population $(\hat{\tau}=1)$ results in full stigma $(S(\hat{\tau})=1)$, which is not crucial for the model prediction.

\begin{figure}[H]
    \centering
    \includegraphics[width=0.75\textwidth]{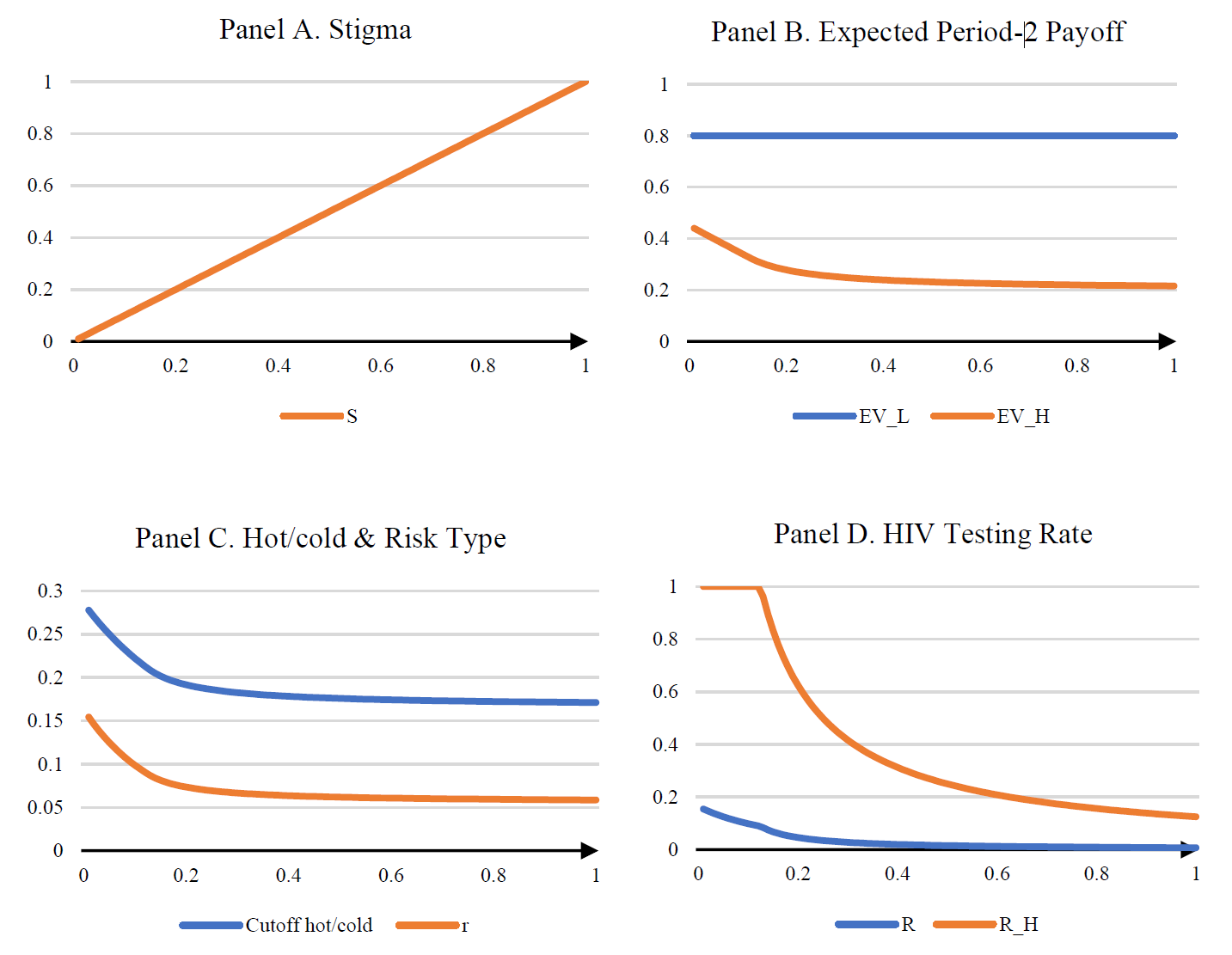}
    \caption{Key Model Parameters in Variance with $t$}
    \label{fig:welfare_parameters}
\end{figure}

Figure \ref{fig:welfare_parameters} shows how each key model parameters vary with different values of the policy instrument, the perceived risk of HIV transmission risk $\hat{\tau}$. As stigma linearly increases with $\hat{\tau}$ (Panel A), the gaps between the expected period-2 payoff of high-risk and low-risk individuals widened (Panel B). As a result, Panel C illustrates the deterrence effect where there are fewer "hot" player $A$ who prefer the "unsafe" equilibrium and, consequently, fewer high-risk players emerge. Panel D illustrates the suppression effect where the testing rates for both the population $\mathcal{A}$ and high-risk player $A$ decrease as stigma is worsened.

\begin{figure}[H]
    \centering
    \includegraphics[width=0.75\textwidth]{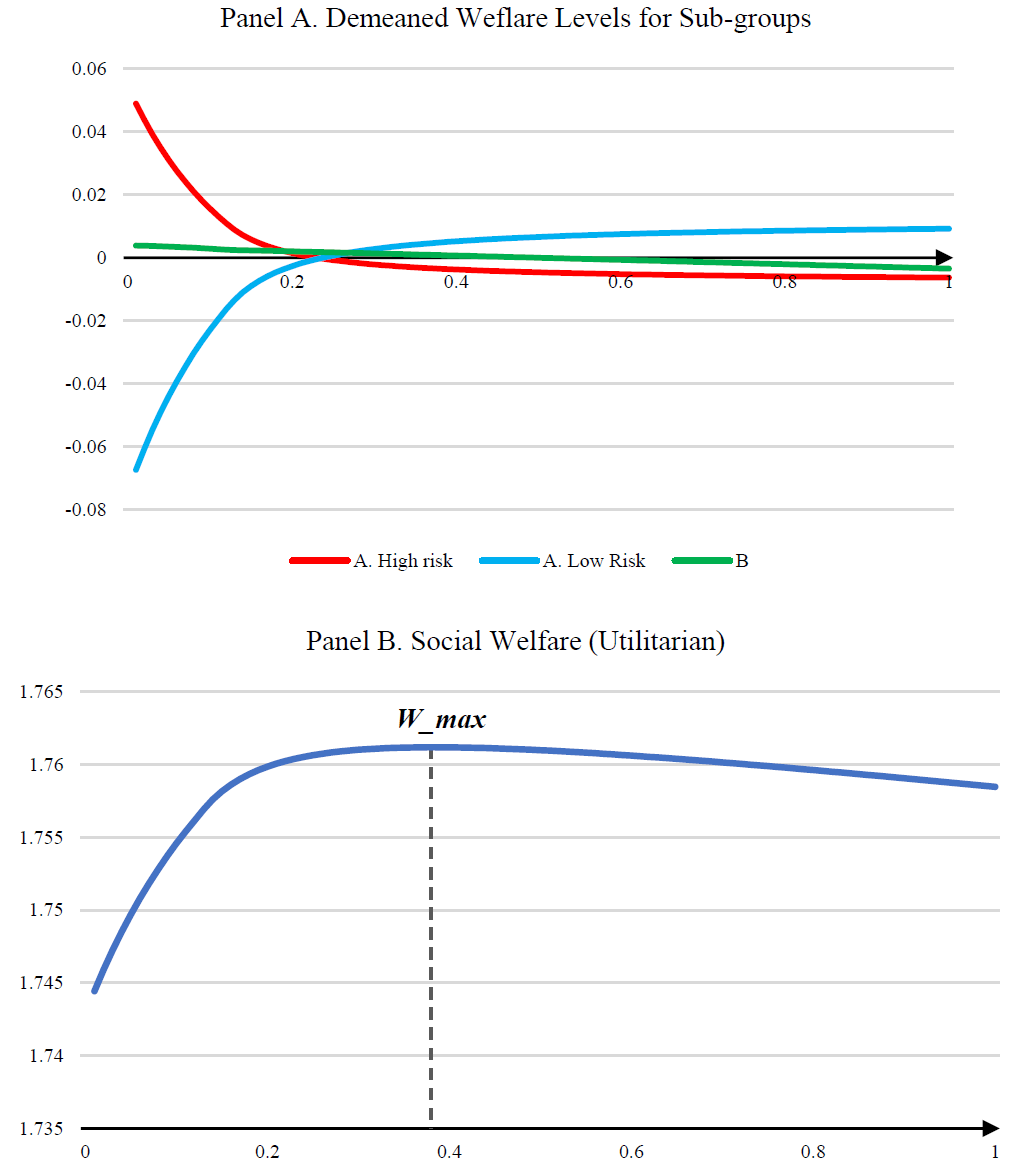}
    \caption{Social Welfare Calculation in Variance with $t$}
    \label{fig:welfare_optimal}
\end{figure}

Figure \ref{fig:welfare_optimal} Panel A presents the demeaned welfare calculation for three types of people, namely high-risk and low-risk player $A$ and player $B$, respectively. The total welfare for each sub-group is calculated as per-capita welfare times the proportion of the sub-group. As $\hat{\tau}$ (or equivalently, stigma $S$) increases, for smaller values, the welfare of low-risk player $A$ increases sharply, which is entirely due to composition change. On the other hand, the sharp decrease in the welfare of high-risk populations is because of both a decrease in per-capita welfare $(M+EV_{H})$ and a reduction in the fraction of high-risk people $r$. In comparison, the welfare of the population $B$ varies minimally throughout because stigmatization $S$ and $E(y_{b}|y_{b} < \hat{\tau}\theta_{H}z)$ are low for small values of $\hat{\tau}$ and testing rate $R$ is low for large values of $\hat{\tau}$.

Figure \ref{fig:welfare_optimal} Panel B presents how the social welfare for the entire population changes as $\hat{\tau}$ varies. When the population is completely misinformed $(\hat{\tau}=1)$, the suppression effect dominates, and the government spending in reducing HIV stigma is welfare-improving. Nevertheless, potentially after some tipping point, the deterrence effect dominates, and reducing stigma is now welfare-reducing even though it greatly improves testing rates. Under the current set of parameter values, a completely un-stigmatized population possesses a lower utilitarian welfare than a completely stigmatized population. The tipping point, the level of stigma that maximizes social welfare, is approximately $\hat{\tau}=0.4$ for the current example. This example shows that the desirability of reducing stigma is not guaranteed and depends on the status quo level of $\hat{\tau}$.

\section{Conclusion}

This study revisits the justification behind the government's effort to reduce HIV stigma. By extending the theoretical framework by Derksen, Muula and van Oosterhout (2020) and Yang et al. (2023) with an additional stage to characterize people's sexual behaviours, the model illustrates the coexistence of suppression and deterrence effect of HIV stigma, which makes its welfare effect ambiguous. A utilitarian welfare analysis confirms that reducing stigma is not always welfare-improving. If the deterrence effect dominates around low levels of stigma, the welfare-maximizing level of stigma will be higher than its natural level. Further reducing perceived transmission risk towards its true value is welfare-reducing.

The current study is an exploratory effort to incorporate the deterrence effect in the welfare evaluation of HIV stigma and it possesses numerous limitations. Firstly, the negative externality of HIV transmission is largely ignored. In the current model, the negative health impact of HIV is limited to single-period one-on-one transmission. Since the model ends at two periods, it ignores the effect of the gradual widespread HIV pandemic and, hence, underestimates the effect of both composition change and testing. Also, and potentially most importantly, the discussion of maintaining (or even increasing) the level of HIV stigma is morally questionable. The core idea of optimizing stigma to achieve welfare improvement is using an error (overestimation of transmission risk) to partially correct another (present bias), yet the former error we intentionally induced and/or maintained causes casualties. In the current model, there are those so present-biased that they would choose an "unsafe" equilibrium regardless of the stigma. These people are strictly worse off when the stigma level increases, which means that they face more severe discrimination, prejudice, and other social disadvantages. This makes the discussion of using them as unsafe-sex-deterrent morally questionable and in need of caution.

\section*{References}

\begin{itemize}
    \item Ariely, D. and Loewenstein, G. (2006) 'The heat of the moment: the effect of sexual arousal on sexual decision making'. Journal of Behavioral Decision Making, 19 (2), pp. 87-98.
    \item Aronson, E. (2011) 'Chapter 5: Self-Justification'. The Social Animal. 11 edn. New York: Worth Publishers, pp. 241.
    \item Bouffard, J.A. (2002) 'The influence of emotion on rational decision making in sexual aggression'. Journal of Criminal Justice, 30 (2), pp. 121-134.
    \item Bouffard, J.A. and Miller, H.A. (2014) 'The Role of Sexual Arousal and Overperception of Sexual Intent Within the Decision to Engage in Sexual Coercion'. Journal of Interpersonal Violence, 29 (11), pp. 1967-1986.
    \item Charness, G. and Rabin, M. (2002) 'Understanding Social Preferences with Simple Tests'. The Quarterly Journal of Economics, 117 (3), pp. 817-869.
    \item Crawford, V.P. (2016) 'New Directions for Modelling Strategic Behavior: Game-Theoretic Models of Communication, Coordination, and Cooperation in Economic Relationships'. Journal of Economic Perspectives, 30 (4), pp. 131-150.
    \item Department of Health and Social Care (UK) (2021) Towards Zero: the HIV Action Plan for England - 2022 to 2025. Available at: https://www.gov.uk/government/publications/towards-zero-the-hiv-action-plan-for-england-2022-to-2025.
    \item Derksen, L., Muula, A. and van Oosterhout, J. (2020) 'Love in the Time of HIV: Testing as a Signal of Risk'. Natural Field Experiments.
    \item Derksen, L., Muula, A. and van Oosterhout, J. (2022) 'Love in the time of HIV: How beliefs about externalities impact health behavior'. Journal of Development Economics, 159 102993.
    \item Farrell, J. (1987) 'Cheap Talk, Coordination, and Entry'. The RAND Journal of Economics, 18 (1), pp. 34-39.
    \item Funk, P. (2004) 'On the effective use of stigma as a crime-deterrent'. European Economic Review, 48 (4), pp. 715-728.
    \item Kahneman, D. (1994) 'New Challenges to the Rationality Assumption'. Journal of Institutional and Theoretical Economics (JITE) / Zeitschrift für die gesamte Staatswissenschaft, 150 (1), pp. 18-36.
    \item Kim, Y.-G. and Sobel, J. (1995) 'An Evolutionary Approach to Pre-Play Communication'. Econometrica, 63 (5), pp. 1181-1193.
    \item Levitt, S.D. and List, J.A. (2007) 'What Do Laboratory Experiments Measuring Social Preferences Reveal About the Real World?". Journal of Economic Perspectives, 21 (2), pp. 153-174.
    \item Major, B. and O'Brien, L.T. (2005) 'The social psychology of stigma'. Annual Review of Psychology, 56 393-421.
    \item O'Donoghue, T. and Rabin, M. (2006) 'Optimal sin taxes'. Journal of Public Economics, 90 (10), pp. 1825-1849.
    \item Rabin, M. (1994) 'A Model of Pre-game Communication'. Journal of Economic Theory, 63 (2), pp. 370-391.
    \item Yang, D. et al. (2023) 'Knowledge, stigma, and HIV testing: An analysis of a widespread HIV/AIDS program'. Journal of Development Economics, 160 102958.
\end{itemize}

\end{document}